# Improving the Performance of TCP over the ATM-UBR service [1 2]

Rohit Goyal, Raj Jain, Shiv Kalyanaraman, Sonia Fahmy, Bobby Vandalore
Department of Computer and Information Science,
395 Dreese Lab, 2015 Neil Avenue
The Ohio State University
Columbus, OH 43210-1277
E-mail: goyal@cis.ohio-state.edu
Phone: (614)-688-4482. Fax: (614)-292-2911

**Abstract**

In this paper we study the design issues for improving TCP performance over the ATM UBR service. ATM-UBR switches respond to congestion by dropping cells when their buffers become full. TCP connections running over UBR can experience low throughput and high unfairness. Intelligent switch drop policies and end-system policies can improve the performance of TCP over UBR with limited buffers. We describe the various design options available to the network as well as to the end systems to improve TCP performance over UBR. We study the effects of Early Packet Discard, and present a per-VC accounting based buffer management policy. We analyze the performance of the buffer management policies with various TCP end system congestion control policies including slow start and congestion avoidance, fast retransmit and recovery and selective acknowledgments. We present simulation results for various small and large latency configurations with varying buffer sizes and number of sources.

## 1 Introduction

The Unspecified Bit Rate (UBR) service provided by ATM networks has no explicit congestion control mechanisms [20]. However, it is expected that many TCP implementations will use the UBR service category. TCP employs a window based end-to-end congestion control mechanism to recover from segment loss and avoids congestion collapse. Several studies have analyzed the performance of TCP over the UBR service. TCP sources running over ATM switches with limited buffers experience low throughput and high unfairness [3, 5, 16, 17].

Figure 1 illustrates a framework for the various design options available to networks and end-systems for congestion control. Intelligent drop policies at switches can be used to improve throughput of transport connections. Early Packet Discard (EPD) [19] has been shown to improve TCP throughput but not fairness [5]. Enhancements that perform intelligent cell drop policies at the switches need to be developed for UBR to improve transport layer throughput and fairness. A policy for selective cell drop based on per-VC buffer management can be used to improve fairness. Providing guaranteed minimum rate to the UBR traffic has also been discussed as a possible candidate to improve TCP performance over UBR.

In addition to network based drop policies, end-to-end flow control and congestion control policies can be effective in improving TCP performance over UBR. The fast retransmit and recovery mechanism [6], can be used in addition to slow start and congestion avoidance to quickly recover from isolated segment losses. The selective acknowledgments (SACK) option has been proposed to recover quickly from multiple segment losses. A change to TCP's fast retransmit and recovery has also been suggested in [2] and [9].

In this paper, we propose a per-VC buffer management scheme called Selective Drop to improve TCP performance over UBR. This scheme is simpler than the Fair Buffer Allocation (FBA) scheme proposed in [8]. We present an analysis of the operation of these schemes and the effects of their parameters. We also provide guidelines for choosing the best parameters for FBA and Selective Drop. We then present simulation results for TCP over the various UBR enhancements. Although, other buffer management schemes have been presented in recent literature, their performance has not been evaluated with the various TCP congestion control options, and for different latencies. We evaluate the performance of the enhancements to UBR, as well as to TCP congestion control mechanisms. We study the performance and interoperability of the network and the end-system enhancements for low latency and large latency configurations.

[1] To appear in Computer Communications

[2] This research was partially sponsored by the NASA Lewis Research Center under contract number NAS3-97198

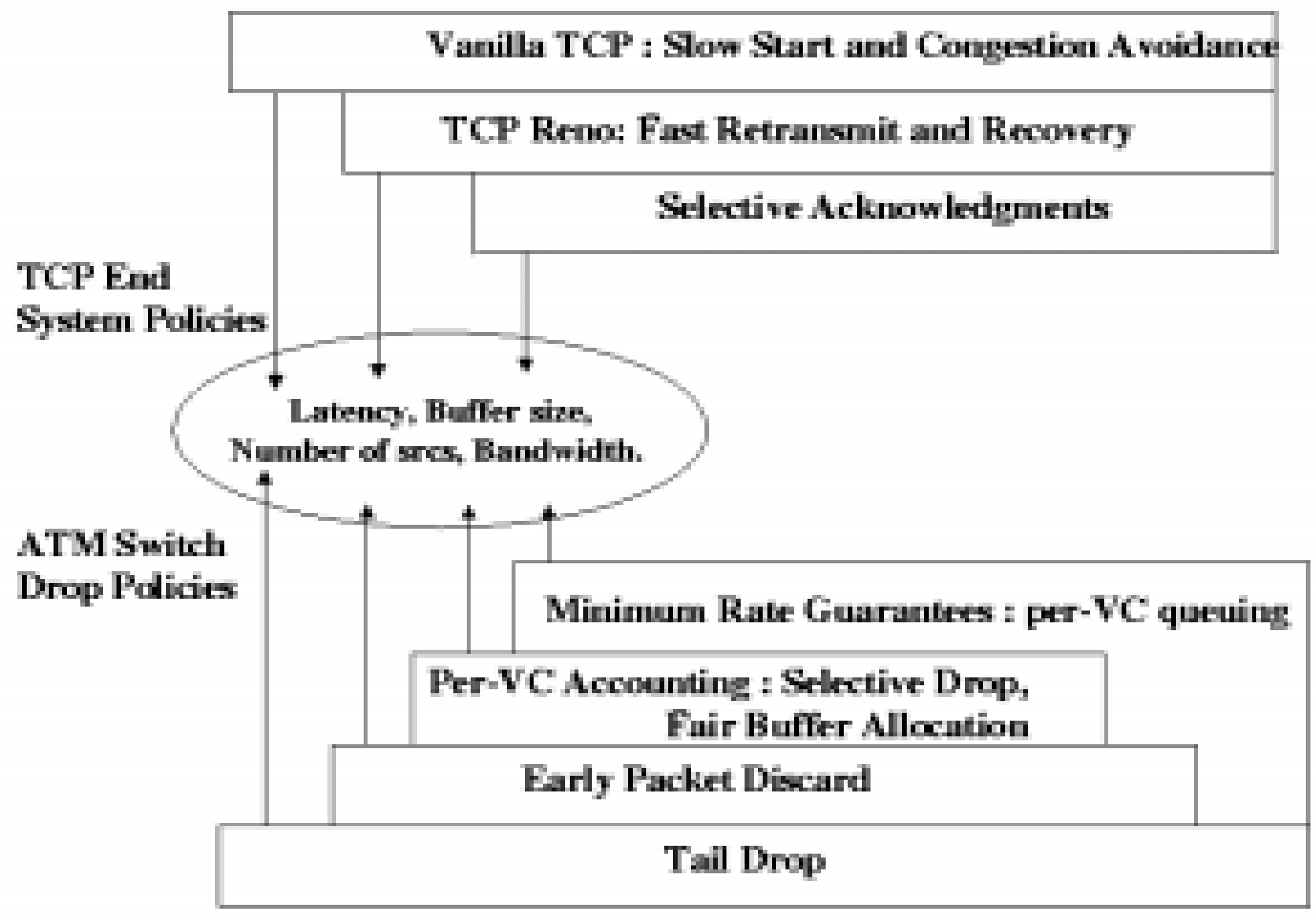

Figure 1: Design Issues for TCP over UBR

We first discuss the congestion control mechanisms in the TCP protocol and explain why these mechanisms can result in low throughput during congestion. We then describe our simulation setup used for all our experiments and define our performance metrics. We present the performance of TCP over vanilla UBR and explain why TCP over vanilla UBR results in poor performance. Section 4 describes in detail, the enhancements to the UBR service category and our implementations. These enhancements include EPD, Selective Drop and Fair Buffer Allocation. Section 5 describes the enhancements to TCP including fast retransmit and recovery, New Reno, and Selective Acknowledgments. We describe our implementations, and present our simulation results for the TCP modifications with each of the UBR changes. Section 7 presents a summary and ideas for future work.

## 2 TCP congestion control (Vanilla TCP)

TCP uses a window based protocol for flow control. TCP connections provide end-to-end flow control to limit the number of packets in the network. The flow control is enforced by two windows. The receiver's window (RCVWND) is enforced by the receiver as measure of its buffering capacity. The congestion window (CWND) is kept at the sender as a measure of the capacity of the network. The sender sends data one window at a time, and cannot send more than the minimum of RCVWND and CWND into the network.

The basic TCP congestion control scheme (we will refer to this as vanilla TCP) consists of the "Slow Start" and "Congestion Avoidance" phases. The variable SSTHRESH is maintained at the source to distinguish between the two phases. The source starts transmission in the slow start phase by sending one segment (typically 512 Bytes) of data, i.e., CWND = 1 TCP segment. When the source receives an acknowledgment for a new segment, the source increments CWND by 1. Since the time between the sending of a segment and the receipt of its ack is an indication of the Round Trip Time (RTT) of the connection, CWND is doubled every round trip time during the slow start phase. The slow start phase continues until CWND reaches SSTHRESH (typically initialized to 64K bytes) and then the congestion avoidance phase begins. During the congestion avoidance phase, the source increases its CWND by 1/CWND every time a segment is acknowledged. The slow start and the congestion avoidance phases correspond to an exponential increase and a linear increase of the congestion window every round trip time respectively.

If a TCP connection loses a packet, the destination responds by sending duplicate acks for each out-of-order packet received. The source maintains a retransmission timeout for the last unacknowledged packet. The timeout value is reset each time a new segment is acknowledged. The source detects congestion by the triggering of the retransmission timeout. At this point, the source sets SSTHRESH to half of CWND. More precisely, SSTHRESH is set to max{2, min{CWND/2, RCVWND}}. CWND is set to one segment size.

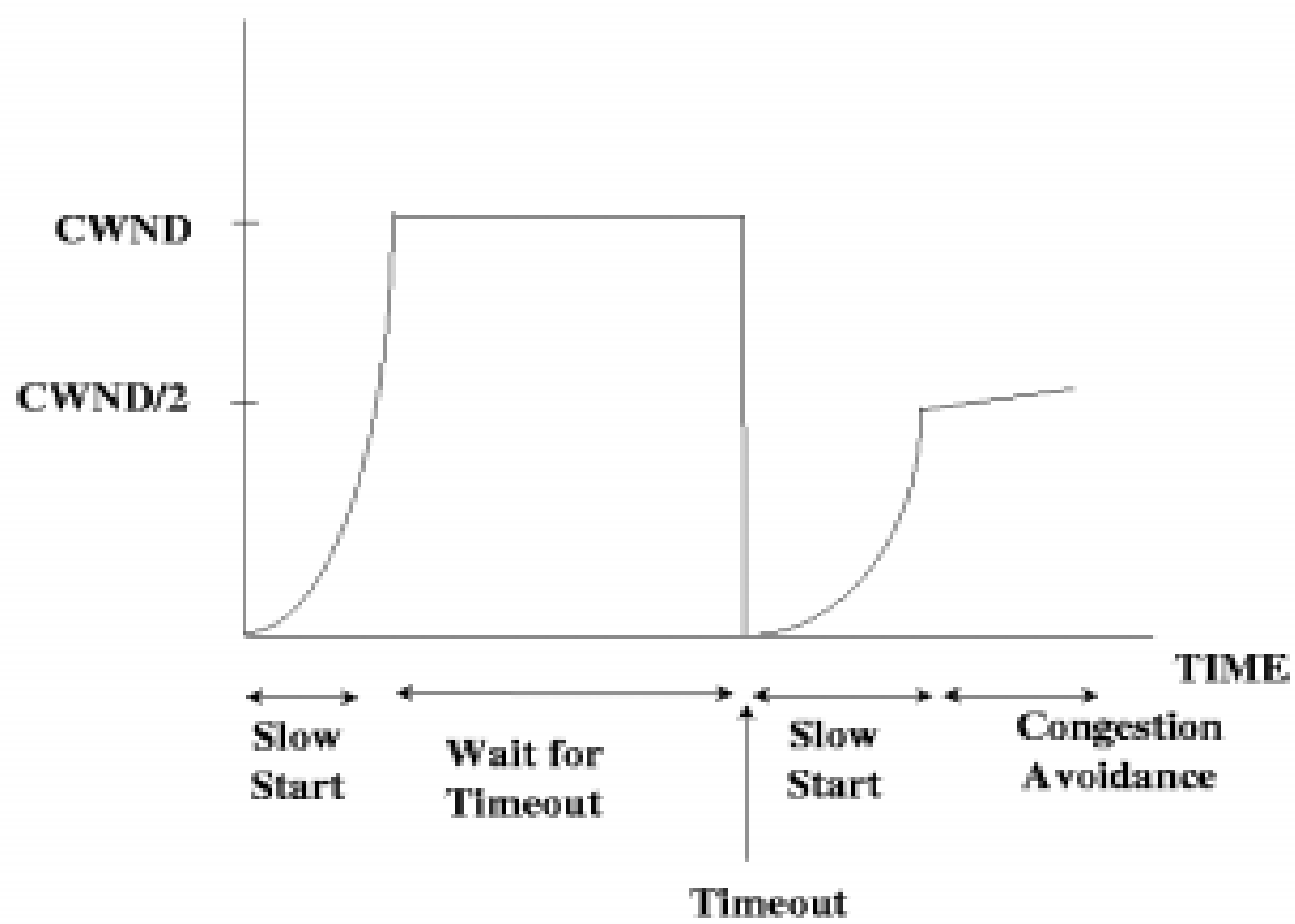

Figure 2: TCP CWND vs Time

As a result, CWND < SSTHRESH and the source enters the slow start phase. The source then retransmits the lost segment and increases its CWND by one every time a new segment is acknowledged. It takes $log_2(\text{CWND}_{\text{orig}}/(2 \times \text{MSS}))$ RTTs from the point when the congestion was detected, for CWND to reach the target value of half its original size ($\text{CWND}_{\text{orig}}$). Here MSS is the TCP maximum segment size value in bytes. This behavior is unaffected by the number of segments lost from a particular window.

If a single segment is lost, and if the receiver buffers out of order segments, then the sender receives a cumulative acknowledgment and recovers from the congestion. Otherwise, the sender attempts to retransmit all the segments since the lost segment. In either case, the sender congestion window increases by one segment for each acknowledgment received, and not for the number of segments acknowledged. The recovery behavior corresponds to a go-back-N retransmission policy at the sender. Note that although the congestion window may increase beyond the advertised receiver window (RCVWND), the source window is limited by the minimum of the two. The typical changes in the source window plotted against time are shown in Figure 2.

Most TCP implementations use a 500 ms timer granularity for the retransmission timeout. The TCP source estimates the Round Trip Time (RTT) of the connection by measuring the time (number of ticks of the timer) between the sending of a segment and the receipt of the ack for the segment. The retransmission timer is calculated as a function of the estimates of the average and mean-deviation of the RTT [10]. Because of coarse grained TCP timers, when there is loss due to congestion, significant time may be lost waiting for the retransmission timeout to trigger. Once the source has sent out all the segments allowed by its window, it does not send any new segments when duplicate acks are being received. When the retransmission timeout triggers, the connection enters the slow start phase. As a result, the link may remain idle for a long time and experience low utilization.

**Coarse granularity TCP timers and retransmission of segments by the go-back-N policy are the main reasons that TCP sources can experience low throughput and high file transfer delays during congestion.**

# 3 TCP over UBR

In its simplest form, an ATM switch implements a tail drop policy for the UBR service category. When a cell arrives at the FIFO queue, if the queue is full, the cell is dropped, otherwise the cell is accepted. If a cell is dropped, the TCP source loses time waiting for the retransmission timeout. Even though TCP congestion mechanisms effectively recover from loss, the resulting throughput can be very low. It is also known that simple FIFO buffering with tail drop results in excessive wasted bandwidth. Simple tail drop of ATM cells results in the receipt of incomplete segments. When part of a segment is dropped at the switch, the incomplete segment is dropped at the destination during reassembly. This wasted bandwidth further reduces the effective TCP throughput. In this section we describe our simulation results to exhibit the poor performance of TCP over UBR. We first describe our simulation model and performance metrics and then go on to discuss our simulation results.

## 3.1 Simulation Model

All simulations presented in this paper are performed on the N source configuration shown in Figure 3. The configuration consists of N identical TCP sources that send data whenever allowed by the window. The switches implement UBR service with optional drop policies described in this paper. The following simulation parameters are used [1]:

- The configuration consists of N identical TCP sources as shown in Figure 3.
- All sources are infinite TCP sources. The TCP layer always sends a segment as long as it is permitted by the TCP window.
- All link delays are 5 microseconds for LANs and 5 milliseconds for WANs[3]. Thus, the Round Trip Time due to the propagation delay is 30 microseconds and 30 milliseconds for LAN and WAN respectively.
- All link bandwidths are 155.52 Mbps.
- Peak Cell Rate is 155.52 Mbps.

[3] In this paper, we refer to low latency connections as LAN connections and high latency connections as WAN connections. LAN and WAN do not refer to the legacy LAN/WAN architectures.

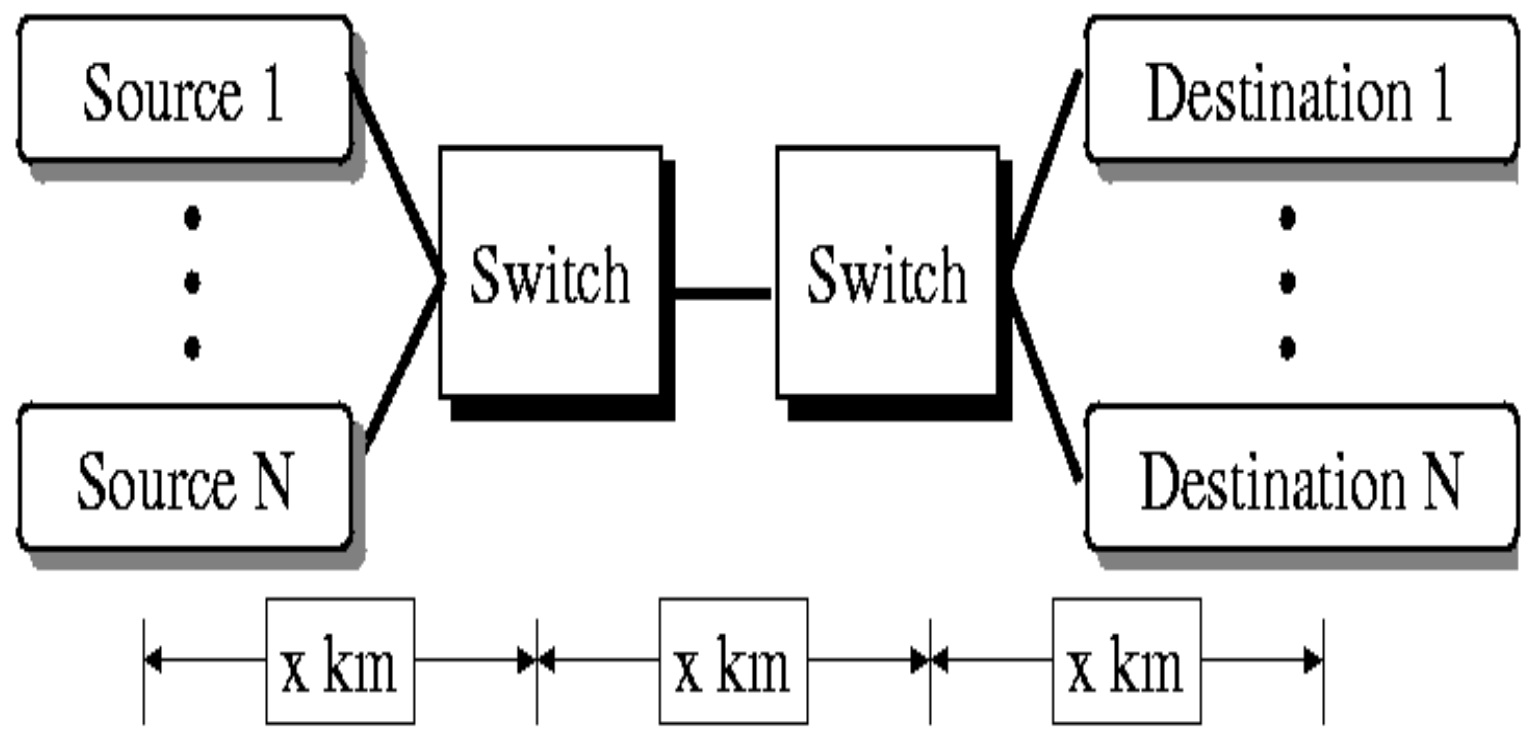

Figure 3: The N-source TCP configuration

- The traffic is unidirectional. Only the sources send data. The destinations send only acknowledgments.
- The TCP segment size is set to 512 bytes. This is the standard value used by many current TCP implementations. Some TCP implementations are using larger values of TCP segment sizes. The effect of segment size on TCP throughput is beyond the scope of this study.
- TCP timer granularity is set to 100 ms. This affects the triggering of retransmission timeout due to packet loss. The values used in most TCP implementations is 500 ms, and some implementations use 100 ms. Several other studies have used smaller TCP timer granularity and have obtained higher throughput values. However, the timer granularity is an important factor in determining the amount of time lost during congestion. Small granularity results in less time being lost waiting for the retransmission timeout to trigger. This results in faster recovery and higher throughput. However, TCP implementations do not use timer granularities of less than 100 ms, and producing results with lower granularity artificially increases the throughput. Moreover, the TCP RTT measurement algorithm is suited for coarse granularity timers. Moderate variation in RTT with a fine granularity timer can result in false TCP timeouts.
- TCP maximum receiver window size is 64K bytes for LANs. This is the default value used in TCP. For WANs, this value is not enough to fill up the pipe, and reach full throughput. In the WAN simulations we use the TCP window scaling option to scale the window to the bandwidth delay product of approximately 1 RTT. The window size used for WANs is 600000 Bytes.
- TCP delay ack timer is NOT set. Segments are acked as soon are they are received.
- Duration of simulation runs is 10 seconds for LANs and 20 seconds for WANs.
- All TCP sources start sending at the same time, and continue to send data through the duration of the simulation. This increases the probability of synchronization among connections, and represents a worst case scenario.

### 3.2 Performance Metrics

The performance of TCP over UBR is measured by the efficiency and fairness which are defined as follows. Let $x_i$ be the throughput of the $i$th TCP source ($0 < i < N$, where $N$ is the total number of TCP sources). Let $C$ be the maximum TCP throughput achievable on the link. Let $E$ be the efficiency of the network. Then, $E$ is defined as

$$E = \frac{\sum_{i=1}^{i=N} x_i}{C}$$

The TCP throughputs $x_i$s are measured at the destination TCP layers. Throughput is defined as the total number of bytes delivered to the destination application divided by the total simulation time. The results are reported in Mbps.

The maximum possible TCP throughput $C$ is the throughput attainable by the TCP layer running over UBR on a 155.52 Mbps link. For 512 bytes of data (TCP maximum segment size), the ATM layer receives 512 bytes of data + 20 bytes of TCP header + 20 bytes of IP header + 8 bytes of LLC header + 8 bytes of AAL5 trailer. These are padded to produce 12 ATM cells. Thus, each TCP segment results in 636 bytes at the ATM Layer. From this, the maximum possible throughput = 512/636 = 80.5% = 125.2 Mbps approximately on a 155.52 Mbps link. For ATM over SONET, this number is further reduced to 120.5 Mbps.

Fairness is measured by the Fairness Index $F$ defined by:

$$\text{Fairness Index (F)} \;=\; \frac{(\sum_{i=1}^{1=N} x_i/e_i)^2}{N \times \sum_{i=1}^{i=N} (x_i/e_i)^2}$$

where $e_i$ is the expected value (fair share) of the throughput for source $i$. For the n-source configuration, $e_i$ is simply an equal share of the total link capacity. Thus, the fairness index metric applies well to our n-source symmetrical configurations. In general, for a more complex complex configuration, the value of $e_i$ can be derived from a more rigorous formulation of a fairness definition that provides max-min fairness to the connections. Note that when $x_1 = x_2 = \ldots = x_n$ then fairness index = 1. Also, low values of the fairness index represent high unfairness among the connections. The desired values of the fairness index must be close to 1. We consider a fairness index of 0.99 to be near perfect. A fairness index of 0.9 may or may not be acceptable depending on the application and the number of sources involved. Details on the fairness metric can be found in [13].

### 3.3 TCP over UBR: Simulation Results

We simulated 5 and 15 TCP sources with finite buffered switches. The simulations were performed with two values of switch buffer sizes both for LAN and WAN links. For WAN experiments, we chose buffer sizes of approximately 1 and 3 times the round trip bandwidth-delay product of the connection. Thus, we selected WAN buffer sizes of 12000 and 36000 cells. For LANs, 1 round trip $\times$ bandwidth is a very small number (11 cells) and is not practical as the size for the buffer. For LAN links, the buffer sizes chosen were 1000 and 3000 cells. These numbers are closer to the buffer sizes of current LAN switches and have been used by other studies on TCP over UBR [3, 16, 17]. The values for WANs were chosen in multiples of round trip time because most ABR feedback control mechanisms can achieve good steady state performance in a fixed number of round trip times, and have similar buffer requirements for zero loss at the switch. Studies on TCP over ABR have used similar values of buffer sizes for both LANs and WANs [14]. It is interesting to assess the performance of TCP over UBR in this situation.

Table 1: TCP over UBR : UBR enhancements (Efficiency)

| Config-uration | Number of Sources | Buffer Size (cells) | UBR | EPD | Selective Drop | FBA |
|---|---|---|---|---|---|---|
| LAN | 5 | 1000 | 0.21 | 0.49 | 0.75 | 0.88 |
| LAN | 5 | 3000 | 0.47 | 0.72 | 0.90 | 0.92 |
| LAN | 15 | 1000 | 0.22 | 0.55 | 0.76 | 0.91 |
| LAN | 15 | 3000 | 0.47 | 0.91 | 0.94 | 0.95 |
| **Column Average** | | | 0.34 | 0.67 | 0.84 | 0.92 |
| WAN | 5 | 12000 | 0.86 | 0.90 | 0.90 | 0.95 |
| WAN | 5 | 36000 | 0.91 | 0.81 | 0.81 | 0.81 |
| WAN | 15 | 12000 | 0.96 | 0.92 | 0.94 | 0.95 |
| WAN | 15 | 36000 | 0.92 | 0.96 | 0.96 | 0.95 |
| **Column Average** | | | 0.91 | 0.90 | 0.90 | 0.92 |

Column 4 of tables 1 and 2 show the efficiency and fairness values respectively for these experiments. Several observations can be made from these results.

- **TCP over vanilla UBR results in low fairness in both LAN and WAN configurations.** This is due to TCP synchronization effects. TCP connections are synchronized when their sources timeout and retransmit at the same time. This occurs because packets from all sources are dropped forcing them to enter the slow start phase. However, in this case, when the switch buffer is about to overflow, one or two connections get lucky and their entire windows are accepted while the segments from all other connections are dropped. All these connections wait for a timeout and stop sending data into the network. The connections that were not dropped send their next window and keep filling up the buffer. All other connections timeout and retransmit at the same time. This results in their segments being dropped again and the synchronization effect is seen. The sources that escape the synchronization get most of the bandwidth. The synchronization effect is particularly important when the number of competing connections is small.
- **The default TCP maximum window size leads to low efficiency in LANs.** LAN simulations have very low efficiency values (less than 50%) while WAN simulations have higher efficiency values. For LANs, the the TCP receiver window size (65535 Bytes) corresponds to more than 1500 cells at the switch for each source. For 5 sources and a buffer size of 1000 cells, the sum of the window sizes is almost 8 times the buffer size. For WAN simulations, with 5 sources and a buffer size of 12000 cells, the sum of the window sizes is less than 6 times the buffer size. As a result, the WAN simulations have higher throughputs than LANs. For LAN experiments with smaller window sizes (less than the default), higher efficiency values are seen.

**Buffer Requirements**

TCP performs best when there is zero loss. In this situation, TCP is able to fill the pipe and fully utilize the link bandwidth. During the exponential rise phase (slow start), TCP sources send out two segments for every segment that is acked. For N TCP sources, in the worst case, a switch can receive a whole window's worth of segments from N-1 sources while it is still clearing out segments from the window of the Nth source. As a result, the switch can have buffer occupancies of up to the sum of all the TCP maximum sender window sizes.

Table 3 contains the simulation results for TCP running over the UBR service with infinite buffering. The maximum queue length numbers give an indication of the buffer sizes required at the switch to achieve zero loss for TCP. The connections achieve 100% of the possible throughput and perfect fairness. For the five source LAN configuration, the maximum queue length is 7591 cells = 7591 / 12 segments = 633 segments $\approx$ 323883 Bytes. This is approximately equal to the sum of the TCP window sizes (65535 $\times$ 5 bytes = 327675 bytes). For the five source WAN configuration, the maximum queue length is 59211 cells = 2526336 Bytes. This is slightly less that the sum of the TCP window sizes (600000 $\times$ 5 = 3000000 Bytes). This is because the switch has 1 RTT to clear out almost 500000 bytes of TCP data (at 155.52 Mbps) before it receives the next window of data. In any case, the increase in buffer requirements is proportional to the number of sources in the simulation. The maximum queue is reached just when the TCP connections reach the maximum window. After that, the window stabilizes and TCP's self clocking congestion mechanism puts one segment into the network for each segment that leaves the network. **For a switch to guarantee zero loss for TCP over UBR, the amount of buffering required is equal to the sum of the TCP maximum window sizes for all the TCP connections.** Note that the maximum window size is determined by the minimum of the sender's congestion window and the receiver's window.

Table 2: TCP over UBR : UBR enhancements (Fairness)

| Configuration | Number of Sources | Buffer Size (cells) | UBR | EPD | Selective Drop | FBA |
|---|---|---|---|---|---|---|
| LAN | 5 | 1000 | 0.68 | 0.57 | 0.99 | 0.98 |
| LAN | 5 | 3000 | 0.97 | 0.84 | 0.99 | 0.97 |
| LAN | 15 | 1000 | 0.31 | 0.56 | 0.76 | 0.97 |
| LAN | 15 | 3000 | 0.80 | 0.78 | 0.94 | 0.93 |
| **Column Average** | | | 0.69 | 0.69 | 0.92 | 0.96 |
| WAN | 5 | 12000 | 0.75 | 0.94 | 0.95 | 0.94 |
| WAN | 5 | 36000 | 0.86 | 1 | 1 | 1 |
| WAN | 15 | 12000 | 0.67 | 0.93 | 0.91 | 0.97 |
| WAN | 15 | 36000 | 0.77 | 0.91 | 0.89 | 0.97 |
| **Column Average** | | | 0.76 | 0.95 | 0.94 | 0.97 |

Table 3: TCP over UBR: Buffer requirements for zero loss

| Number of Sources | Configuration | Efficiency | Fairness | Maximum Queue (Cells) |
|---|---|---|---|---|
| 5 | LAN | 1 | 1 | 7591 |
| 15 | LAN | 1 | 1 | 22831 |
| 5 | WAN | 1 | 1 | 59211 |
| 15 | WAN | 1 | 1 | 196203 |

For smaller buffer sizes, efficiency typically increases with increasing buffer sizes (see table 1). Larger buffer sizes result in more cells being accepted before loss occurs, and therefore higher efficiency. This is a direct result of the dependence of the buffer requirements to the sum of the TCP window sizes. The buffer sizes used in the LAN simulations reflect the typical buffer sizes used by other studies of TCP over ATM ([3, 14, 16]), and implemented in ATM workgroup switches.

# 4 UBR Enhancements

From the simulation results in section 3.3, it is clear that TCP over UBR can experience poor performance. In this section, we study enhancements to the UBR service category that improve TCP performance over UBR.

## 4.1 Early Packet Discard

The Early Packet Discard (EPD) policy [19] has been suggested to remedy some of the problems with tail drop switches. EPD drops complete packets instead of partial packets. As a result, the link does not carry incomplete packets which would have been discarded during reassembly. A threshold R less than the buffer size, is set at the switches. When the switch queue length exceeds this threshold, all cells from any new packets are dropped. Packets which had been partly received before exceeding the threshold are still accepted if there is buffer space. In the worst case, the switch could have received one cell from all N connections before its buffer exceeded the threshold. To accept all the incomplete packets, there should be additional buffer capacity of upto the sum of the packet sizes of all the connections. Typically, the threshold R should be set to the buffer size − N × the maximum packet size, where N is the expected number of connections active at any time.

The EPD algorithm used in our simulations is the one suggested by [16, 17]. Column 5 of tables 1 and 2 show the efficiency and fairness respectively of TCP over UBR with EPD. The switch thresholds are selected so as to allow one entire packet from each connection to arrive after the threshold is exceeded. We use thresholds of Buffer Size − 200 cells in our simulations. 200 cells are enough to hold one packet each from all 15 TCP connections. This reflects the worst case scenario when all the fifteen connections have received the first cell of their packet and then the buffer occupancy exceeds the threshold.

Tables 1 and 2 show that **EPD improves the efficiency of TCP over UBR, but it does not significantly improve fairness.** This is because EPD indiscriminately discards complete packets from all connections without taking into account their current rates or buffer utilizations. When the buffer occupancy exceeds the threshold, all new packets are dropped. There is a more significant improvement in fairness for WANs because of the relatively larger buffer sizes.

## 4.2 Per-VC Buffer Management

Intelligent buffer management schemes have been proposed that use per-VC accounting to maintain the current buffer utilization of each UBR VC. Fair Buffer Allocation [8] and Selective Drop are two such schemes. In these schemes, a fair allocation is calculated for each VC, and if the VC's buffer occupancy exceeds its fair allocation, its next incoming packet is dropped. Both schemes maintain a threshold $R$, as a fraction of the buffer capacity $K$. When the total buffer occupancy exceeds R×K, new packets are dropped depending on the VC's (say $VC_i$) buffer occupancy ($Y_i$).

Selective Drop keeps track of the activity of each VC by counting the number of cells from each VC in the buffer. A VC is said to be active if it has at least one cell in the buffer. A fair allocation is calculated as the current buffer occupancy divided by number of active VCs.

Let the buffer occupancy be denoted by $X$, and the number of active VCs be denoted by $N_a$. Then the fair allocation or fair share $Fs$ for each VC is given by,

$$Fs = \frac{X}{N_a}$$

The ratio of the number of cells of a VC in the buffer to the fair allocation gives a measure of how much the VC is overloading the buffer i.e., by what ratio it exceeds the fair allocation. Let $Y_i$ be the number of cells from $VC_i$ in the buffer. Then the Load Ratio, $L_i$, of $VC_i$ is defined as

$$L_i = \frac{Y_i}{Fs}$$

or

$$L_i = \frac{Y_i \times N_a}{X}$$

If the load ratio of a VC is greater than a parameter $Z$, then new packets from that VC are dropped in preference to packets of a VC with load ratio less than $Z$. Thus, $Z$ is used as a cutoff for the load ratio to indicate that the VC is overloading the switch.

Figure 4 illustrates the drop conditions for Selective Drop. For a given buffer size $K$ (cells), the selective drop scheme assigns a static minimum threshold parameter $R$ (cells). If the buffer occupancy $X$ is less than or equal to this minimum threshold $R$, then no cells are dropped. If the buffer occupancy is greater than $R$, then the next new incoming packet of $VC_i$ is dropped if the load ratio of $VC_i$ is greater than $Z$.

We performed simulations to find the value of $Z$ that optimizes the efficiency and fairness values. We first performed 5 source LAN simulations with 1000 cell buffers. We set $R$ to 0.9 $\times$ the buffer size $K$. This ensured that there was enough buffer space to accept incomplete packets during congestion. We experimented with values of $Z$ = 2, 1, 0.9, 0.5 and 0.2. $Z = 0.9$ resulted in good performance. Further simulations of values of $Z$ around 0.9 showed that $Z = 0.8$ produces the best efficiency and fairness values for this configuration. For WAN simulations, any $Z$ value between 0.8 and 1 produced the best results.

The Fair Buffer Allocation Scheme proposed by [8] uses a more complex form of the parameter Z and compares it with the load ratio $L_i$ of a VC. To make the cutoff smooth, FBA uses the current load level in the switch. The scheme compares the load ratio of a VC to another threshold that determines how much the switch is congested. For a given buffer size $K$, the FBA scheme assigns a static Minimum Threshold parameter $R$ (cells). If the buffer occupancy $X$ is less than or equal to this minimum threshold $R$, then no cells are dropped. When the buffer occupancy is greater than $R$, then upon the arrival of every new packet, the load ratio of the VC (to which the packet belongs) is compared to an allowable drop threshold $T$ calculated as

$$T = Z \times \frac{K - R}{X - R}$$

In this equation $Z$ is a linear scaling factor. The next packet from $VC_i$ is dropped if

$$(X > R) \text{ AND } \frac{Y_i \times N_a}{X} > Z \times \frac{K - R}{X - R}$$

Figure 4 shows the switch buffer with buffer occupancies $X$ relative to the minimum threshold $R$ and the buffer size $K$ where incoming TCP packets may be dropped.

Note that when the current buffer occupancy $X$ exceeds the minimum threshold $R$, it is not always the case that a new packet is dropped. The load ratio in the above equation determines if $VC_i$ is using more than a fair amount of buffer space. $X/N_a$ is used as a measure of a fair allocation for each VC, and $Z \times ((K - R)/(X - R))$ is a drop threshold for the buffer. If the current buffer occupancy ($Y_i$ is greater than this dynamic threshold times the fair allocation ($X/N_a$), then the new packet of that VC is dropped.

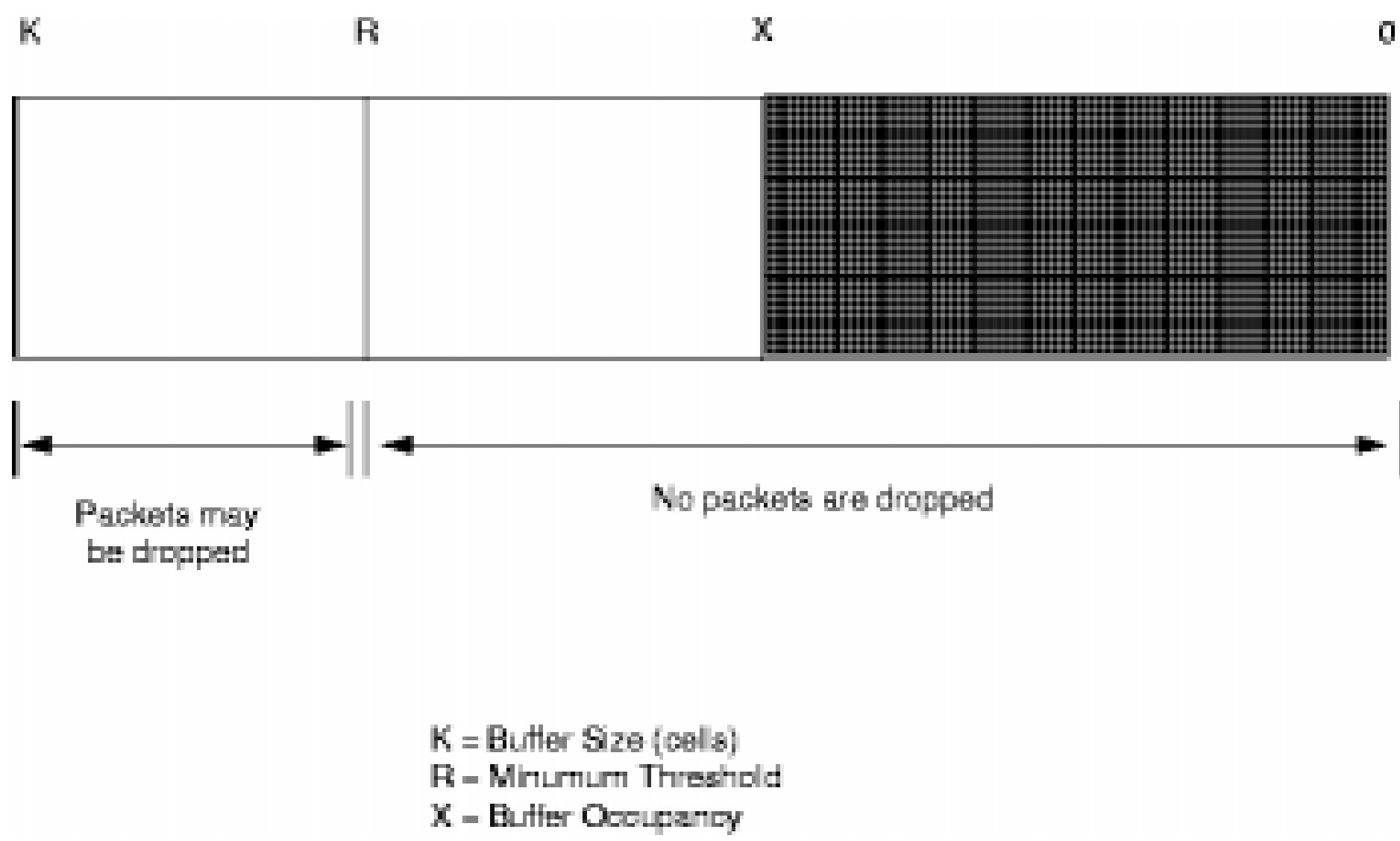

Figure 4: Selective Drop and FBA: Buffer Occupancy for drop

### 4.2.1 Effect of Parameters

**Effect of the minimum drop threshold $R$**

The load ratio threshold for dropping a complete packet is $Z((K - R)/(X - R))$. As $R$ increases for a fixed value of the buffer occupancy $X$, $X - R$ decreases, which means that the drop threshold $((K - R)/(X - R))$ increases and each connection is allowed to have more cells in the buffer. Higher values of $R$ provide higher efficiency by allowing higher buffer utilization. Lower values of $R$ should provide better fairness than higher values by dropping packets earlier.

**Effect of the linear scale factor $Z$**

The parameter $Z$ scales the FBA drop threshold by a multiplicative factor. $Z$ has a linear effect on the drop threshold, where lower values of $Z$ lower the threshold and vice versa. Higher values of $Z$ should increase the efficiency of the connections. However, if $Z$ is very close to 1, then cells from a connection may not be dropped until the buffer overflows.

We performed a full factorial experiment [13] with the following parameter variations for both LANs and WANs. Each experiment was performed for N-source configuration.

- Number of sources, N = 5 and 15.
- Buffer capacity, K = 1000 , 2000 and 3000 cells for LANs and 12000, 24000 and 36000 cells for WANs.
- Minimum drop threshold, R = 0.9×K , 0.5×K and 0.1×K.

- Linear scale factor, Z = 0.2 , 0.5 and 0.8.

A set of 54 experiments were conducted to determine the values of $R$ and $Z$ that maximized efficiency and fairness among the TCP sources. We sorted the results with respect to the efficiency and fairness values. The following observations can be made from the simulation results (see [6]).

- **There is a tradeoff between efficiency and fairness.** The highest values of fairness (close to 1) have the lowest values of efficiency. The simulation data shows that these results are for low R and Z values. Higher values of the minimum threshold R combined with low Z values lead to slightly higher efficiency. Efficiency is high for high values of R and Z. Lower efficiency values have either R or Z low, and higher efficiency values have either of R or Z high. When R is low (0.1), the scheme can drop packets when the buffer occupancy exceeds a small fraction of the capacity. When Z is low, a small rise in the load ratio will result in its packets being dropped. This improves the fairness of the scheme, but decreases the efficiency especially if R is also low. **For configurations simulated, we found that the best value of R was about 0.9 and Z about 0.8.**
- **The fairness of the scheme is sensitive to parameters.** The simulation results showed that small changes in the values of $R$ and $Z$ can result in significant differences in the fairness results. With the increase of $R$ and $Z$, efficiency shows an increasing trend. However there is considerable variation in the fairness numbers. We attribute this to TCP synchronization effects. Sometimes, a single TCP source can get lucky and its packets are accepted while all other connections are dropped. When the source finally exceeds its fair-share and should be dropped, the buffer is no longer above the threshold because all other sources have stopped sending packets and are waiting for timeout.
- **Both Selective Drop and FBA improve both fairness and efficiency of TCP over UBR.** This is because cells from overloading connections are dropped in preference to underloading ones. As a result, Selective Drop is more effective in breaking TCP synchronization. When the buffer exceeds the threshold, only cells from overloading connections are dropped. This frees up some bandwidth and allows the underloading connections to increase their window and obtain more throughput. In general, the average efficiency and fairness values for FBA (for optimal parameter values) are higher than the previously discussed options. Columns 6 and 7 of tables 1,2 show the fairness and efficiency values for Selective Drop and FBA with $R = 0.9$ and $Z = 0.8$ respectively.
- **Fairness and efficiency increase with increase in buffer size.** This supports the discussion in section 3.3 and shows that the performance improves with increasing buffer size for FBA and Selective Drop.

# 5 TCP Enhancements

## 5.1 TCP Reno: Fast Retransmit and Recovery

Current TCP implementations use a coarse granularity (typically 500 ms) timer for the retransmission timeout. As a result, during congestion, the TCP connection can lose much time waiting for the timeout. In Figure 2, the horizontal CWND line shows the time lost in waiting for a timeout to occur. During this time, the TCP neither sends new packets nor retransmits lost packets. Moreover, once the timeout occurs, the CWND is set to 1 segment, and the connection takes several round trips to efficiently utilize the network. TCP Reno implements the fast retransmit and recovery algorithms that enable the connection to quickly recover from isolated segment losses [21].

When a TCP receives an out-of-order segment, it immediately sends a duplicate acknowledgment to the sender. When the sender receives three duplicate ACKs, it concludes that the segment indicated by the ACKs has been lost, and immediately retransmits the lost segment. The sender then reduces its CWND by half (plus 3 segments) and also saves half the original CWND value in SSTHRESH. Now for each subsequent duplicate ACK, the sender inflates CWND by one and tries to send a new segment. Effectively, the sender waits for half a round trip before sending one segment for each subsequent duplicate ACK it receives. As a result, the sender maintains the network pipe at half of its capacity at the time of fast retransmit. This is called "Fast Retransmit."

Approximately one round trip after the missing segment is retransmitted, its ACK is received (assuming the retransmitted segment was not lost). At this time, instead of setting CWND to one segment and proceeding to do slow start, the TCP sets CWND to SSTHRESH, and then does congestion avoidance. This is called "Fast Recovery."

When a single segment is lost from a window, Reno TCP recovers within approximately one RTT of knowing about the loss or two RTTs after the lost packet was first sent. The sender receives three duplicate ACKS one RTT after the dropped packet was sent. It then retransmits the lost packet. For the next round trip, the sender receives duplicate ACKs for the whole window of packets sent after the lost packet. The sender waits for half the window and then transmits a half window worth of new packets. All of this takes one RTT after which the sender receives a new ACK acknowledging the retransmitted packet and the entire window sent before the retransmission. CWND is set to half its original value and congestion avoidance is performed.

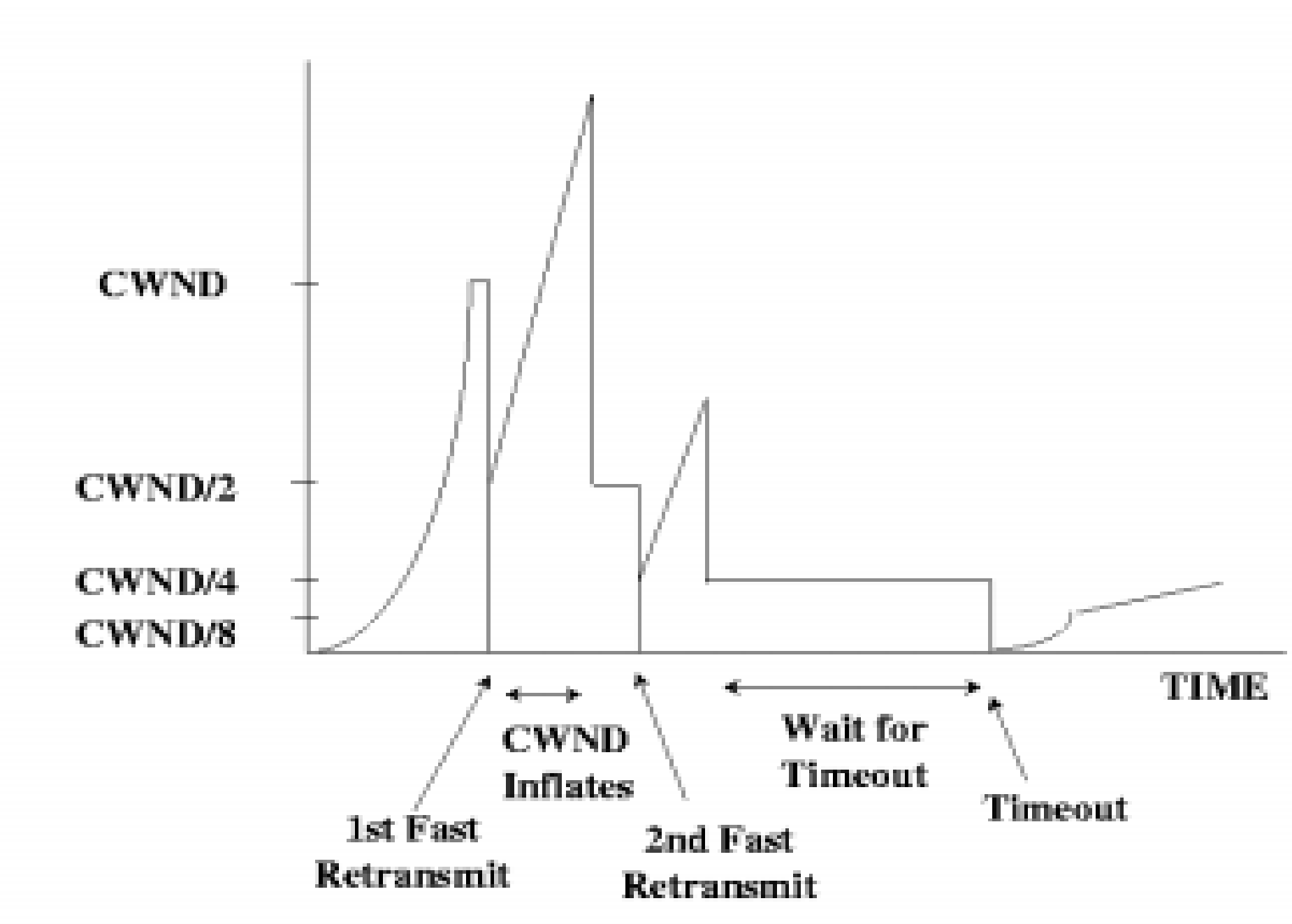

Figure 5: TCP Fast Retransmit and Recovery

## 5.2 TCP Reno: Simulation Results

Tables 4 and 5 list the simulation results of TCP Reno with each of the UBR options. Tables 8 and 9 compare the average efficiency and fairness values with the vanilla TCP results.

- **For long latency connections (WAN), fast retransmit and recovery hurts the efficiency.** This is because congestion typically results in multiple packets being dropped. Fast retransmit and recovery cannot recover from multiple packet losses and slow start is triggered. The additional segments sent by fast retransmit and recovery (while duplicate ACKs are being received) may be retransmitted during slow start. In WAN links with large bandwidth delay products, the number of retransmitted segments can be significant. Thus, fast retransmit can add to the congestion and reduce throughput. Figure 5 shows a case when three consecutive packets are lost from a window, the sender TCP incurs fast retransmit twice and then times out. At that time, SSTHRESH is set to one-eighth of the original congestion window value (CWND in the figure) As a result, the exponential phase lasts a very short time, and the linear increase begins at a very small window. Thus, the TCP sends at a very low rate and loses much throughput.

- **Fast retransmit and recovery improves the efficiency of TCP over UBR for the LAN configuration.** From table 4, the effect of multiple packet losses is much less visible in low latency connections because for a small RTT and large bandwidth, the linear increase very quickly fills up the network pipe. As a result it results in almost same efficiency as the exponential increase.

- **The addition of EPD with fast retransmit and recovery results in a large improvement in both fairness for LANs.** Thus, the combination on EPD and fast retransmit can provide high throughput and fairness for low latency configurations.

Table 4: Reno TCP over UBR (Efficiency)

| Configuration | Number of Sources | Buffer Size (cells) | UBR | EPD | Selective Drop | FBA |
|---|---|---|---|---|---|---|
| LAN | 5 | 1000 | 0.53 | 0.97 | 0.97 | 0.97 |
| LAN | 5 | 3000 | 0.89 | 0.97 | 0.97 | 0.97 |
| LAN | 15 | 1000 | 0.42 | 0.97 | 0.97 | 0.97 |
| LAN | 15 | 3000 | 0.92 | 0.97 | 0.97 | 0.97 |
| **Column Average** | | | 0.69 | 0.97 | 0.97 | 0.97 |
| WAN | 5 | 12000 | 0.61 | 0.79 | 0.8 | 0.76 |
| WAN | 5 | 36000 | 0.66 | 0.75 | 0.77 | 0.78 |
| WAN | 15 | 12000 | 0.88 | 0.95 | 0.79 | 0.79 |
| WAN | 15 | 36000 | 0.96 | 0.96 | 0.86 | 0.89 |
| **Column Average** | | | 0.78 | 0.86 | 0.81 | 0.81 |

## 5.3 TCP New Reno: A Modification to Fast Retransmit and Recovery

In the previous section, we showed that fast retransmit and recovery cannot effectively recover from multiple packet losses. A modification to Reno is proposed in [2, 9] to overcome this shortcoming.

The "fast-retransmit phase" is introduced, in which the sender remembers the highest sequence number sent (RECOVER) when the fast retransmit is first triggered. After the first unacknowledged packet is retransmitted (when three duplicate ACKs are received), the sender follows the usual fast recovery algorithm and inflates the CWND by one for each duplicate ACK it receives. When the sender receives an acknowledgment for the retransmitted packet, it checks if the ACK acknowledges all segments including RECOVER. If so, the ACK is a new ACK, and the sender exits the fast retransmit-recovery phase, sets its CWND to SSTHRESH and starts a linear increase (congestion avoidance). If on the other hand, the ACK is a partial ACK, i.e., it acknowledges the retransmitted segment, and only a part of the segments before RECOVER, then the sender immediately retransmits the next expected segment as indicated by the ACK. This continues until all segments including RECOVER are acknowledged. **This mechanism ensures that the sender will recover from N segment losses in N round trips.** This is called "New Reno."

Table 5: Reno TCP over UBR (Fairness)

| Configuration | Number of Sources | Buffer Size (cells) | UBR | EPD | Selective Drop | FBA |
|---|---|---|---|---|---|---|
| LAN | 5 | 1000 | 0.77 | 0.99 | 0.99 | 0.97 |
| LAN | 5 | 3000 | 0.93 | 0.99 | 1 | 0.99 |
| LAN | 15 | 1000 | 0.26 | 0.96 | 0.99 | 0.69 |
| LAN | 15 | 3000 | 0.87 | 0.99 | 0.99 | 1 |
| **Column Average** | | | 0.71 | 0.98 | 0.99 | 0.91 |
| WAN | 5 | 12000 | 0.99 | 1 | 0.99 | 1 |
| WAN | 5 | 36000 | 0.97 | 0.99 | 0.99 | 1 |
| WAN | 15 | 12000 | 0.91 | 0.96 | 0.99 | 0.95 |
| WAN | 15 | 36000 | 0.74 | 0.91 | 0.98 | 0.98 |
| **Column Average** | | | 0.90 | 0.97 | 0.99 | 0.98 |

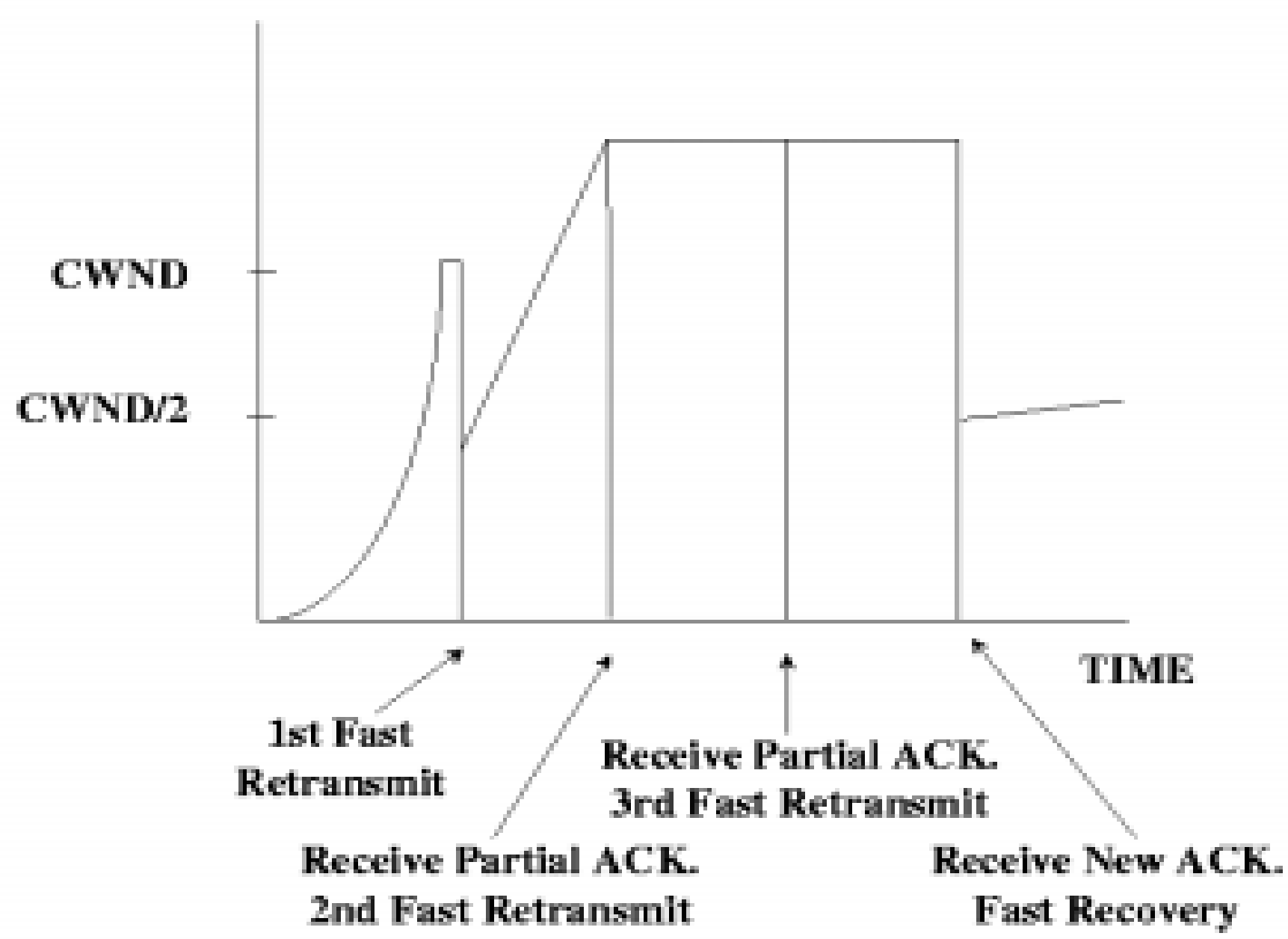

Figure 6: TCP with the fast retransmit phase

As a result, the sender can recover from multiple packet losses without having to time out. In case of small propagation delays, and coarse timer granularities, this mechanism can effectively improve TCP throughput over vanilla TCP. Figure 6 shows the congestion window graph of a TCP connection for three contiguous segment losses. The TCP retransmits one segment every round trip time (shown by the CWND going down to 1 segment) until a new ACK is received.

In our implementation, we combined "New Reno" and SACK TCP as described in the following subsection.

## 5.4 SACK TCP: Selective Acknowledgments

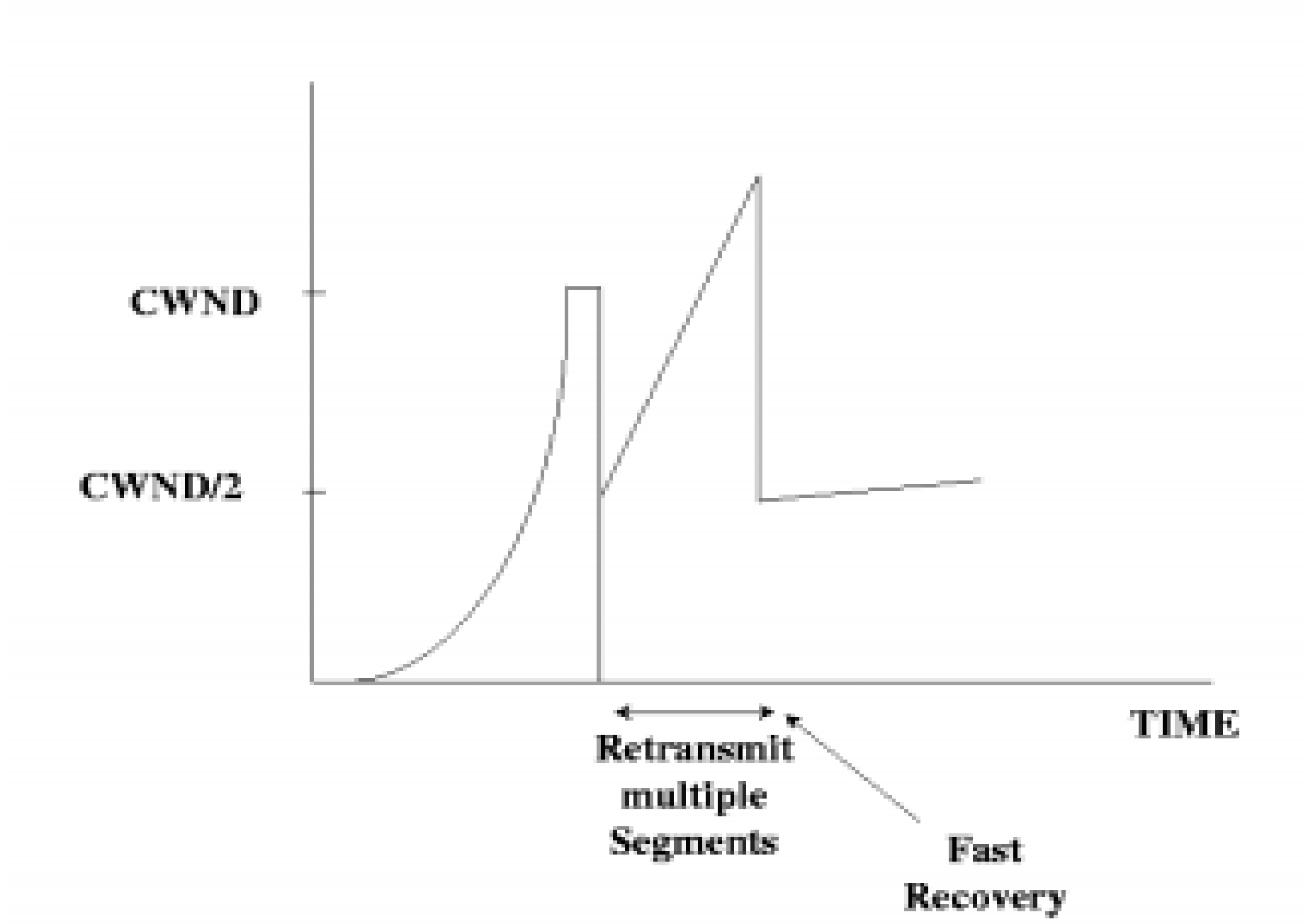

Figure 7: SACK TCP Recovery from packet loss

TCP with Selective Acknowledgments (SACK TCP) has been proposed to efficiently recover from multiple segment losses [18]. In SACK TCP, acknowledgments contain additional information about the segments have been received by the destination. When the destination receives out-of-order segments, it sends duplicate ACKs (SACKs) acknowledging the out-of-order segments it has received. From these SACKs, the sending TCP can reconstruct information about the segments not received at the destination. When the sender receives three duplicate ACKs, it retransmits the first lost segment, and inflates its CWND by one for each duplicate ACK it receives. This behavior is the same as Reno TCP. However, when the sender, in response to duplicate ACKs, is allowed by the window to send a segment, it uses the SACK information to retransmit lost segments before sending new segments. As a result, the sender can recover from multiple dropped segments in about one round trip. Figure 7 shows the congestion window graph of a SACK TCP recovering from segment losses. During the time when the congestion window is inflating (after fast retransmit has incurred), the TCP is sending missing packets before any new packets.

### 5.4.1 SACK TCP Implementation

In this subsection, we describe our implementation of SACK TCP and some properties of SACK. Our implementation is based on the SACK implementation described in [2, 4, 18].

The SACK option is negotiated in the SYN segments during TCP connection establishment. The SACK information is sent with an ACK by the data receiver to the data sender to inform the sender of the out-of-sequence segments received. The format of the SACK packet has been proposed in [18]. The SACK option is sent whenever out of sequence data is received. All duplicate ACK's contain the SACK option. The option contains a list of some of the contiguous blocks of data already received by the receiver. Each data block is identified by the sequence number of the first byte in the block (the left edge of the block), and the sequence number of the byte immediately after the last byte of the block. Because of the limit on the maximum TCP header size, at most three SACK blocks can be specified in one SACK packet.

The receiver keeps track of all the out-of-sequence data blocks received. When the receiver generates a SACK, the first SACK block specifies the block of data formed by the most recently received data segment. This ensures that the receiver provides the most up to date information to the sender. After the first SACK block, the remaining blocks can be filled in any order.

The sender keeps a table of all the segments sent but not ACKed. When a segment is sent, it is entered into the table. When the sender receives an ACK with the SACK option, it marks all the segments specified in the SACK option blocks as SACKed. The entries for each segment remain in the table until the segment is ACKed. The remaining behavior of the sender is very similar to Reno implementations with the modification suggested in Section 5.3 [4]. When the sender receives three duplicate ACKs, it retransmits the first unacknowledged packet. During the fast retransmit phase, when the sender is sending one segment for each duplicate ACK received, it first tries to retransmit the holes in the SACK blocks before sending any new segments. When the sender retransmits a segment, it marks the segment as retransmitted in the table. If a retransmitted segment is lost, the sender times out and performs slow start. When a timeout occurs, the sender resets the SACK table.

During the fast retransmit phase, the sender maintains a variable PIPE that indicates how many bytes are currently in the network pipe. When the third duplicate ACK is received, PIPE is set to the value of CWND and CWND is reduced by half. For every subsequent duplicate ACK received, PIPE is decremented by one segment because the ACK denotes a packet leaving the pipe. The sender sends data (new or retransmitted) only when PIPE is less than CWND. This implementation is equivalent to inflating the CWND by one segment for every duplicate ACK and sending segments if the number of unacknowledged bytes is less than the congestion window value.

When a segment is sent, PIPE is incremented by one. When a partial ACK is received, PIPE is decremented by two. The first decrement is because the partial ACK represents a retransmitted segment leaving the pipe. The second decrement is done because the original segment that was lost, and had not been accounted for, is now actually considered to be lost.

#### 5.4.2 SACK TCP: Recovery Behavior

We now calculate a bound for the recovery behavior of SACK TCP, and show that SACK TCP can recover from multiple packet losses more efficiently than Reno or vanilla TCP. Suppose that at the instant when the sender learns of the first packet loss (from three duplicate ACKs), the value of the congestion window is CWND. Thus, the sender has CWND bytes of data waiting to be acknowledged. Suppose also that the network has dropped a block of data which is CWND/n bytes long (This will typically result in several segments being lost). After one RTT of sending the first dropped segment, the sender receives three duplicate ACKs for this segment. It retransmits the segment, and sets PIPE to CWND $-$ 3, and sets CWND to CWND/2. For each duplicate ACK received, PIPE is decremented by 1. When PIPE reaches CWND, then for each subsequent duplicate ACK received, another segment can be sent. All the ACKs from the previous window take 1 RTT to return. For half RTT nothing is sent (since PIPE $>$ CWND). For the next half RTT, if CWND/n bytes were dropped, then only CWND/2 $-$ CWND/n bytes (of retransmitted or new segments) can be sent. In the second RTT, the sender can retransmit 2(CWND/2 − CWND/n) bytes. This is because for each retransmitted segment in the first RTT, the sender receives a partial ACK that indicates that the next segment is missing. As a result, PIPE is decremented by 2, and the sender can send 2 more segments (both of which could be retransmitted segments) for each partial ACK it receives.

[4] It is not clear to us whether the modification proposed in [9] is necessary with the SACK option. The modification is under further study.

Table 6: SACK TCP over UBR+ : Efficiency

| Configuration | Number of Sources | Buffer (cells) | UBR | EPD | Selective Drop |
|---|---|---|---|---|---|
| LAN | 5 | 1000 | 0.76 | 0.85 | 0.94 |
| LAN | 5 | 3000 | 0.98 | 0.97 | 0.98 |
| LAN | 15 | 1000 | 0.57 | 0.78 | 0.91 |
| LAN | 15 | 3000 | 0.86 | 0.94 | 0.97 |
| **Column Average** | | | 0.79 | 0.89 | 0.95 |
| WAN | 5 | 12000 | 0.90 | 0.88 | 0.95 |
| WAN | 5 | 36000 | 0.97 | 0.99 | 1.00 |
| WAN | 15 | 12000 | 0.93 | 0.80 | 0.88 |
| WAN | 15 | 36000 | 0.95 | 0.95 | 0.98 |
| **Column Average** | | | 0.94 | 0.91 | 0.95 |

Thus, the number of RTTs $N_{rec}$ needed by SACK TCP to recover from a loss of CWND/n is given by

$$N_{rec} \leq log(\frac{n}{n-2}) \text{ for } 2 < n \leq 4$$

If less than one fourth of CWND is lost, then SACK TCP can recover in 1 RTT. If more than one half the CWND is dropped, then there will not be enough duplicate ACKs for PIPE to become large enough to transmit any segments in the first RTT. Only the first dropped segment will be retransmitted on the receipt of the third duplicate ACK. In the second RTT, the ACK for the retransmitted packet will be received. This is a partial ACK and will result in PIPE being decremented by 2 so that 2 packets can be sent. As a result, PIPE will double every RTT, and SACK will recover no slower than slow start [2, 4]. SACK would still be advantageous because timeout would be still avoided unless a retransmitted packet were dropped.

### 5.4.3 SACK TCP: Simulation Results

We performed simulations for the LAN and WAN configurations for three drop policies – tail drop, Early Packet Discard and Selective Drop. Tables 6 and 7 show the efficiency and fairness values of SACK TCP with various UBR drop policies. Tables 8 and 9 show the comparative column averages for Vanilla, Reno and SACK TCP. Several observations can be made from these tables:

- For most cases, for a given drop policy, **SACK TCP provides higher efficiency than the corresponding drop policy in vanilla TCP.** This confirms the intuition provided by the analysis of SACK that SACK recovers at least as fast as slow start when multiple packets are lost. In fact, for most cases, SACK recovers faster than both fast retransmit/recovery and slow start algorithms.
- **For LANs, the effect of drop policies is very important and can dominate the effect of SACK**. For UBR with tail drop, SACK provides a significant improvement over Vanilla and Reno TCPs. However, as the drop policies get more sophisticated, the effect of TCP congestion mechanism is less pronounced. This is because, the typical LAN switch buffer sizes are small compared to the default TCP maximum window of 64K bytes, and so buffer management becomes a very important factor. Moreover, the degraded performance of SACK over Reno in LANs (see tables 8 and 9) is attributed to excessive timeout due to the retransmitted packets being lost. In this case SACK loses several round trips in retransmitting parts of the lost data and then times out. After timeout, much of the data is transmitted again, and this results in wasted throughput. This result reinforces the need for a good switch drop policy for TCP over UBR.

Table 7: SACK TCP over UBR+ : Fairness

| Config-uration | Number of Sources | Buffer (cells) | UBR | EPD | Selective Drop |
|---|---|---|---|---|---|
| LAN | 5 | 1000 | 0.22 | 0.88 | 0.98 |
| LAN | 5 | 3000 | 0.92 | 0.97 | 0.96 |
| LAN | 15 | 1000 | 0.29 | 0.63 | 0.95 |
| LAN | 15 | 3000 | 0.74 | 0.88 | 0.98 |
| **Column Average** | | | 0.54 | 0.84 | 0.97 |
| WAN | 5 | 12000 | 0.96 | 0.98 | 0.95 |
| WAN | 5 | 36000 | 1.00 | 0.94 | 0.99 |
| WAN | 15 | 12000 | 0.99 | 0.99 | 0.99 |
| WAN | 15 | 36000 | 0.98 | 0.98 | 0.96 |
| **Column Average** | | | 0.98 | 0.97 | 0.97 |

- **The throughput improvement provided by SACK is significant for wide area networks**. When the propagation delay is large, a timeout results in the loss of a significant amount of time during slow start from a window of one segment. With Reno TCP (with fast retransmit and recovery), performance is further degraded (for multiple packet losses) because timeout occurs at a much lower window than vanilla TCP. With SACK TCP, a timeout is avoided most of the time, and recovery is complete within a small number of roundtrips. Even if timeout occurs, the recovery is as fast as slow start but some time may be lost in the earlier retransmissions.
- **The performance of SACK TCP can be improved by intelligent drop policies like EPD and Selective drop**. This is consistent with our earlier results with Vanilla and Reno TCP . Thus, we recommend that intelligent drop policies be used in UBR service.
- **The fairness values for selective drop are comparable to the values with the other TCP versions**. Thus, SACK TCP does not hurt the fairness in TCP connections with an intelligent drop policy like selective drop. The fairness of tail drop and EPD are sometimes a little lower for SACK TCP. This is again because retransmitted packets are lost and some connections time out. Connections which do not time out do not have to go through slow start, and can utilize more of the link capacity. The fairness among a set of heterogeneous TCP connections is a topic of further study.

# 6 Effect of a Large Number of Sources

In workgroup and local area networks, the number of TCP connections active at any given time is small, and can be realistically modeled by the above simulation results. However, in wide area networks, more than 15 TCP connections may be simultaneously active. It becomes interesting to assess the effectiveness of Selective Drop to provide high efficiency and fairness to a large number of sources.

Table 8: TCP over UBR: Comparative Efficiencies

| Configuration | UBR | EPD | Selective Drop |
|---|---|---|---|
| | **LAN** | | |
| **Vanilla TCP** | 0.34 | 0.67 | 0.84 |
| **Reno TCP** | 0.69 | 0.97 | 0.97 |
| **SACK TCP** | 0.79 | 0.89 | 0.95 |
| | **WAN** | | |
| **Vanilla TCP** | 0.91 | 0.9 | 0.91 |
| **Reno TCP** | 0.78 | 0.86 | 0.81 |
| **SACK TCP** | 0.94 | 0.91 | 0.95 |

Table 9: TCP over UBR: Comparative Fairness

| Configuration | UBR | EPD | Selective Drop |
|---|---|---|---|
| | **LAN** | | |
| **Vanilla TCP** | 0.69 | 0.69 | 0.92 |
| **Reno TCP** | 0.71 | 0.98 | 0.99 |
| **SACK TCP** | 0.54 | 0.84 | 0.97 |
| | **WAN** | | |
| **Vanilla TCP** | 0.76 | 0.95 | 0.94 |
| **Reno TCP** | 0.90 | 0.97 | 0.99 |
| **SACK TCP** | 0.98 | 0.97 | 0.97 |

Even with a large number of TCP connections, EPD does not significantly affect fairness over vanilla UBR, because EPD does not perform selective discard of packets based on buffer usage. However, with a large number of sources, the fairness metric can take high values even in clearly unfair cases. This is because, as the number of sources increases, the effect of a single source or a few sources on the the fairness metric decreases. As a result, vanilla UBR might have a fairness value of 0.95 or better even if a few TCP's receive almost zero throughput. The effect of unfairness is easily seen with a small number of sources. Vanilla UBR and EPD are clearly unfair, and the performance of Selective Drop needs to be tested with large number of sources for a more strict value of fairness. [13] suggests that a value of 0.99 for the fairness metric reflects high fairness even for a large number of sources. Selective Drop should provide high efficiency and fairness in such cases.

We performed experiments with 50 and 100 TCP sources using SACK TCP and Selective Drop. The experiments were performed for WAN and satellite networks with the N source configuration. The simulations produced efficiency values of 0.98 and greater with fairness values of 0.99 and better for buffer sizes of 1 RTT and 3 RTT. The simulations produced high efficiency and fairness for fixed buffer sizes, irrespective of the number of sources (5, 15, 50 or 100). The details of the simulation results with a large number sources have been published as part of a separate study in [15]. The simulation results illustrate that SACK TCP with a per-VC buffer management policy like Selective Drop can produce high efficiency and fairness even for a large number of TCP sources.

From the simulation and analysis presented in this paper, we know that vanilla TCP performs poorly because TCP sources waste bandwidth when they are waiting for a timeout. Reno TCP performs poorly in the case of multiple packet losses because of timeout, and congestion avoidance at a very low window size. The effect of these behaviors is mitigated with a large number of sources. When a large number of sources are fairly sharing the link capacity, each TCP gets a small fraction of the capacity, and the steady state window sizes of the TCPs are small. When packets are lost from a few TCPs, other TCPs increase their congestion widows to utilize the unused capacity within a few round trips. As a result, overall link efficiency improves, but at the expense of the TCPs suffering loss. The TCPs that lose packets recover the fastest with SACK TCP. Thus, SACK TCP can help in quickly achieving fairness after packet loss.

# 7 Summary

In this paper, we have shown techniques for improving TCP performance over the UBR service category in ATM networks. We summarize the results in the form of a comparative analysis of the various options for TCP and UBR. This summary is based on the choice of optimal parameters for the drop policies. For both selective drop and fair buffer allocation, the values of R and Z are chosen to be 0.9 and 0.8 respectively.

- **To achieve maximum possible throughput (or zero cell loss) for TCP over UBR, switches need buffers equal to the sum of the receiver windows of all the TCP connections.**
- **With limited buffer sizes, TCP performs poorly over vanilla UBR switches.** TCP throughput is low, and there is unfairness among the connections. The coarse granularity TCP timer is an important reason for low TCP throughput.
- **UBR with EPD improves the throughput performance of TCP.** This is because partial packets are not being transmitted by the network and some bandwidth is saved. EPD does not have much effect on fairness because it does not drop segments selectively.

- **UBR with selective packet drop using per-VC accounting improves fairness over UBR+EPD.** Connections with higher buffer occupancies are more likely to be dropped in this scheme. The efficiency values are similar to the ones with EPD.
- **UBR with the Fair Buffer Allocation scheme can improve TCP throughput and fairness.** There is a tradeoff between efficiency and fairness and the scheme is sensitive to parameters. We found R = 0.9 and Z = 0.8 to produce best results for our configurations.
- **Fast retransmit and recovery is detrimental to the performance of TCP over large delay-bandwidth links**. This is because fast retransmit and recovery cannot effectively recover from multiple packet losses.
- **Selective Acknowledgments with TCP further improves the performance of TCP over UBR.** SACK TCP results in better throughput than both vanilla and Reno TCP. The fairness and efficiency also increase with intelligent UBR drop policies.
- **End-to-end policies have a more significant effect on large latency networks while drop policies have more impact on low latency networks.**
- SACK TCP with a per-VC accounting based buffer management policy like Selective Drop can produce high efficiency and fairness for TCP/IP over UBR even for a large number of TCP sources.

To conclude, TCP performance over UBR can be improved by either improving TCP using selective acknowledgments, or by introducing intelligent buffer management policies at the switches. Efficient buffer management has a more significant influence on LANs because of the limited buffer sizes in LAN switches compared to the TCP maximum window size. In WANs, the drop policies have a smaller impact because both the switch buffer sizes and the TCP windows are of the order of the bandwidth-delay product of the network. Also, the TCP linear increase is much slower in WANs than in LANs because the WAN RTTs are higher.

In this paper we have not presented a comprehensive comparative study of TCP performance with other per-VC buffer management schemes. This is a topic of further study, and is beyond the scope of this paper. Also, the present version of Selective Drop assigns equal weight to the competing connections. Selective Drop with weighted fairness is a topic of further study.

[5] All our papers and ATM Forum contributions are available from http://www.cis.ohio-state.edu/~jain